\documentclass[twocolumn,showpacs,preprintnumbers,amsmath,amssymb]{revtex4}
\usepackage{epsfig}

\def\lsim{\mathrel{\rlap{\lower4pt\hbox{\hskip1pt$\sim$}}
    \raise1pt\hbox{$<$}}}                % less than or approx. symbol
\def\gsim{\mathrel{\rlap{\lower4pt\hbox{\hskip1pt$\sim$}}
    \raise1pt\hbox{$>$}}}                % greater than or approx. symbol

\begin{document}

\title{Local Temperature and Universal Heat Conduction in FPU chains}

\author{Trieu Mai$^1$, Abhishek Dhar$^2$, Onuttom Narayan$^1$}
\affiliation{$^1$Department of Physics, University of California, 
Santa Cruz, California 95064, USA}
\affiliation{$^2$Raman Research Institute, Bangalore 560080, India}

\date{\today}

\begin{abstract}
It is shown numerically that for Fermi Pasta Ulam (FPU) chains with
alternating masses and heat baths at slightly different temperatures
at the ends, the local temperature (LT) on small scales behaves
paradoxically in steady state. This expands the long established
problem of equilibration of FPU chains.  A well-behaved LT appears
to be achieved for equal mass chains; the thermal conductivity is
shown to diverge with chain length $N$ as $N^{1/3},$ relevant for
the much debated question of the universality of one dimensional
heat conduction.  The reason why earlier simulations have obtained
systematically higher exponents is explained.  
\end{abstract}


\maketitle 
It has long been established~\cite{FPU} that Fermi Pasta Ulam (FPU)
chains (one dimensional chains of particles with anharmonic forces
between them) may not be able to achieve thermal equilibrium with
seemingly reasonable initial conditions.  This has interesting
implications for the ergodic hypothesis, at the foundation of
statistical mechanics. The results of Ref.~\cite{FPU} led to the
discovery of solitons in continuum versions~\cite{soliton}, and
eventually an understanding of chaotic dynamics.  With regard to
the initial results, a vast body of work has established~\cite{chaos,BGG}
that quasi-periodic solutions exist below an energy threshold, above
which the system does equilibrate.

Although all this work has been with closed boundary conditions,
the study of FPU chains has expanded to include heat bath boundary
conditions: with slightly different temperatures imposed at the two
ends, the thermal conductivity can be measured and shows anomalous
properties~\cite{BLR,LLPreview}.  Since the baths at the ends are
at different temperatures, the concept of equilibrium has to be
extended: a local temperature (LT) that varies smoothly along the
chain has to be defined.  Surprisingly, this has not been fully
investigated, even though the discussion of heat conductivity is
in terms of Fourier's law~\cite{BLR} which is meaningless if the
local temperature is ill-behaved. When the heat baths at the two
ends have {\it equal\/} temperatures, one can prove analytically
that the only possible steady state is the one in thermal
equilibrium~\cite{bergmann55}.

In this paper, we demonstrate for the first time that, for FPU
chains with heat bath boundary conditions, the LT behaves paradoxically
on small scales.  In particular, we show through numerical simulations
that for FPU-$\beta$ chains with alternating light and heavy masses,
connected to heat baths at temperatures $T_L$ and $T_R$ at the left
and right end respectively (with $\Delta T = T_L - T_R$ sufficiently
small that the system is in the linear response regime), the kinetic
temperature of the particles oscillates as one moves along the
chain.  The ratio of amplitude of these oscillations to $\Delta
T/N$ is constant as $\Delta T$ is reduced, and increases with the
chain length $N$: in the vicinity of the $2i$'th particle, the
temperature difference between heavy and light particles scales
approximately as $\delta T\sim [\Delta T/\sqrt N] f(i/N).$ Thus if
one were to coarse grain over a region of the order of the mean
free path, the intra-cell temperature uncertainty $O(\Delta T/\sqrt
N)$ would dominate the $O(\Delta T/N)$ change in temperature between
adjacent cells.  (Similar results are obtained when the heavy and
light particles are randomly ordered, so that {\it no\/} O(1)
coarse-graining length would work.) Nor is this a boundary effect
with a characteristic decay length, since the dependence on $i$ is
through $i/N.$ We also show that there are no problems when the FPU
chain has equal masses.  For one dimensional gases, lack of energy
equipartition between heavy and light particles has been observed
for hard particle systems\cite{lightheavy}, but we have verified
that $T(x)$ as a function of position $x$ (instead of particle
number $i$) is smooth. For the FPU system, where the lattice constant
can be taken to be arbitrarily large, this resolution of the problem
is not applicable.

For equal mass FPU chains, having verified the existence of a
well-behaved LT, we measure the heat conductivity $\kappa(N)$ and
demonstrate that
\begin{equation}
\kappa(N)\sim N^\alpha
\label{defalpha}
\end{equation}
with $\alpha=1/3,$ in agreement with the earlier renormalization
group (RG) analytical result~\cite{RG} for a fluid model.  We do
this by simulating very long FPU chains with up to $N=65536$
particles, showing that $\alpha=1/3$ is attained in this regime.
This result is insensitive to system parameters, in contrast to the
apparent exponents for smaller $N$. We also explain why the apparent
$\alpha$ for smaller $N$ is systematically higher than 1/3, as seen
in various earlier numerical
simulations~\cite{LLPreview,LLP98,LLP03,DLLP1,WangLi}.  This supports
the assertion that there is only one universality class for heat
conduction in one dimensional momentum conserving systems (that
have a well-behaved LT), contrary to earlier
suggestions~\cite{LLP98,LLP03,DLLP1,WangLi}.

The generalized FPU chain consists of a sequence of particles
connected by springs between nearest neighbors. The Hamiltonian is
$\sum_i m_i \dot x_i^2/2 + V(x_i - x_{i+1}),$ where the potential
energy of the interparticle springs is $V(z) =
{k_2}z^2/2+{k_3}z^3/3+{k_4}z^4/4+...$~.  For all figures shown in
this paper, we use the FPU-$\beta$ model where $k_2=k_4=1$ and other
$k_i$'s are zero.  As noted in the original FPU paper~\cite{FPU},
the incidental even symmetry of this potential may cause non-mixing
of even and odd modes, but this is not the case with the heat bath
boundary conditions used here.  After fixing $\Delta T,$ we allow
the system to reach steady state and then measure the kinetic
temperature $T_i = m_i\langle v_i^2\rangle$ of each particle.  Unless
otherwise noted, the end particles $i=1$ and $i=N$ are connected
to Langevin baths at temperatures $T_L=2.0$ and $T_R=0.5$ respectively,
by adding damping and noise terms to their equations of motion,
which are integrated with an accurate Verlet-like
algorithm~\cite{verletbath}.

\begin{figure}
\begin{center}
\includegraphics[width=3in]{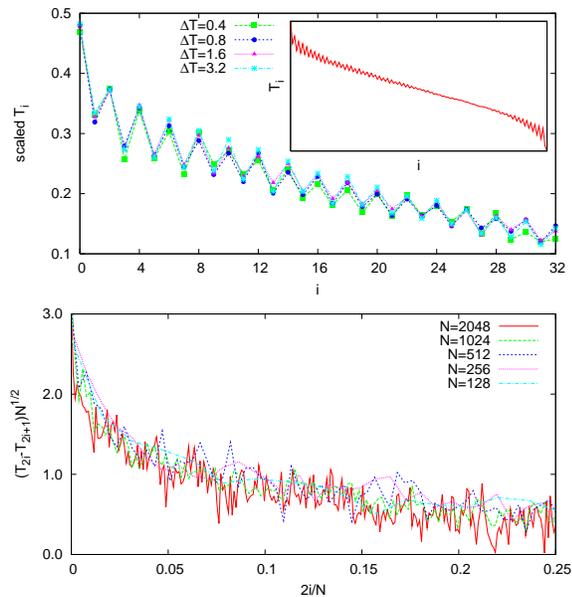}
\caption{(Color online) Oscillations of the kinetic temperature
profile for FPU-$\beta$ dimer chains with mass ratio 2.62, $\gamma
= 2.$ (Top inset) Full profile for a system with $N=128$ and
$T_L=2.0$, $T_R=0.5$.  (Top) $N=128$ particles, $T_{L,R}=2\pm\Delta
T/2.$ The scaled temperature is $(T-2)/\Delta T$; the different
plots line up.  (Bottom) $(T_{2i}-T_{2i+1})N^{1/2}$ versus $2i/N$
for different $N$. $T_{L,R}=2.0,0.5.$ The $N^{1/2}$ and $2i/N$ are
chosen to approximately match the vertical and horizontal scales
of the plots.  Together, the figures imply $\delta T \approx [\Delta
T/\sqrt N] f(i/N)$.  }
\label{fig:alt}
\end{center}
\end{figure}

Fig.\ref{fig:alt} shows $T_i$ as a function of particle number $i$
for a dimer chain with $N=128$ and mass ratio 2.62. An unusual
oscillating temperature profile is seen; the lighter particles are
hotter than the heavier ones on the left and colder on the right.
Since the sign, not just the magnitude, of $T_{i+1} - T_i$ oscillates,
it is not because of a thermal conductivity $\kappa$ that oscillates
on the microscopic scale.  The behavior shown here is robust to
changes in the mass ratio, although the oscillation amplitude
changes, or the interparticle potential (e.g. if an exponential
potential is used, i.e. a Toda lattice with alternating masses).
Oscillations in $T_i$ as a function of $i$ are also seen when the
Langevin baths are replaced with Nose Hoover baths~\cite{NoseHoover}.

Fig.\ref{fig:alt} also shows that if $\Delta T$ is reduced or $N$
is increased, the temperature difference between the $2i$'th and
$(2i + 1)$'th particles scales approximately as $\delta T\sim [\Delta
T/\sqrt N] f(i/N).$ Thus coarse graining over a region of size
$O(1),$ comparable to the mean free path, creates an unusual LT:
the uncertainty in temperature in a coarse-grained region is greater
than the variation between adjacent regions.  Moreover, the `decay
length' over which the oscillations penetrate into the interior of
the chain is a fixed fraction of $N,$ showing that this is not
negligible for large $N.$ Non-monotonic temperature profiles are
also seen when the heavy and light particles are ordered randomly,
in which case the heavy and light kinetic temperatures track two
separate smooth curves.

Our results show that anharmonicity and disorder are not sufficient
for a well-behaved LT even with heat bath boundaries imposing a
$O(1/N)$ temperature gradient.  When a well-behaved LT {\it is\/}
achieved, e.g. (as we will show) for equal mass FPU chains, it
should be viewed as being fragile.  The question remains: what are
the necessary and sufficient conditions for a local temperature?

\begin{figure}
\begin{center}
\includegraphics[width=3in]{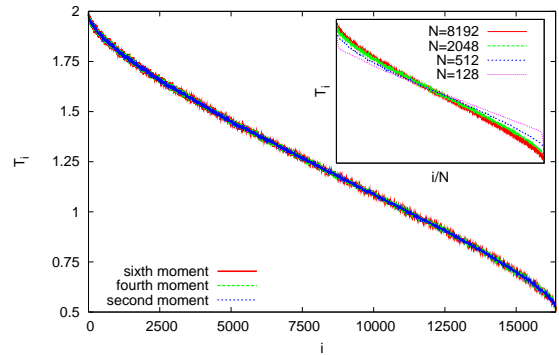}
\caption{(Color online) Kinetic temperature profile for a FPU-$\beta$
chain with $N=16384$, $T_L=2.0$, $T_R=0.5$ and $\gamma=2.0$.  The
first three even moments of the velocity are shown; their agreement
indicates a Gaussian velocity distribution.  (Inset) Normalized
temperature profiles for different  $N$.}
\label{fig:temp}
\end{center}
\end{figure}

We also simulate FPU-$\beta$ chains of equal mass particles (of
unit mass) as described in the previous paragraphs.  When steady
state is reached, we observe an approximately linear temperature
profile as shown by Fig.(\ref{fig:temp}).
There is a slight curvature near the boundaries, which decreases
with $N$.  Fig.(\ref{fig:temp}) also shows that the velocity
distributions are Gaussian (at least for large $N$) which is necessary
for a LT. This indicates that the equal mass FPU chain with heat
baths at the boundaries has a well-behaved LT.

With this reassurance, we proceed to measure the heat conductivity
as a function of $N$ for the equal mass case.  With a small temperature
difference applied across the system, Fourier's law predicts that
the heat current $j$ should be equal to $-\kappa \nabla T,$ with a
$\kappa$ that depends on microscopic properties.  Equivalently, $j
= -\kappa (\Delta T/N),$ where we have defined $\nabla T$ as $\Delta
T/N$. By measuring $j(N),$ deviations from $j\sim 1/N$ are interpreted
as a $N$-dependent conductivity $\kappa (N)$ and a consequent
breakdown of Fourier's law.

For one dimensional momentum conserving systems where a well-behaved
LT exists, a RG study of the hydrodynamic equations of a normal
fluid~\cite{RG} showed that Eq.(\ref{defalpha}) is satisfied with
$\alpha = 1/3.$ This has been confirmed by simulations of hard
particle gases~\cite{Grassberger,Cipriani,RC}, although very large
systems are required~\cite{foot:smallrc} and the issue is not
completely settled~\cite{gas25}. On the other hand, numerical
simulations of oscillator chains, including FPU chains, give various
exponents~\cite{LLPreview,LLP98,LLP03,DLLP1,WangLi} for different
systems, often slightly higher than 1/3. This seems consistent with
early results from mode-coupling theory (MCT), which predict a heat
conductivity exponent of $\alpha=2/5$~\cite{LLP98,LLP03,WangLi},
although recent MCT analyses predict exponents that depend on the
leading nonlinearity~\cite{DLLP1,DLLP2} and the extent of transverse
motion~\cite{WangLi}. The apparent agreement between the numerical
and MCT results has led to speculation that there may be two (or
more) universality classes with different exponents~\cite{WangLi,DLLP1}.
The most recent MCT analysis~\cite{DLLP1,DLLP2} predicts that
$\alpha\neq 1/3$ is restricted to even potentials $V(z)$, which is
why we have studied the FPU-$\beta$ model here.

Recently, two of us have extended the earlier RG treatment to systems
with broken symmetry~\cite{tunable}, which can occur on intermediate
length scales in one dimension (as in nanotubes). The resulting
crystalline hydrodynamic equations also yielded $\alpha=1/3,$ like
the earlier fluid result~\cite{foot:fluid}.  From numerical results,
it was argued that the apparent $\alpha > 1/3$ found for FPU chains
is probably a crossover effect from hard particle systems. However,
the possibility that the crossover could be pushed out to
$N\rightarrow\infty$ could not be ruled out. Thus the numerical and
analytical evidence so far allow for FPU chains to be a singular
limit for the RG with a special value of the conductivity exponent
$\alpha,$ motivating our numerical simulations.

\begin{figure}
\begin{center}
\includegraphics[width=3in]{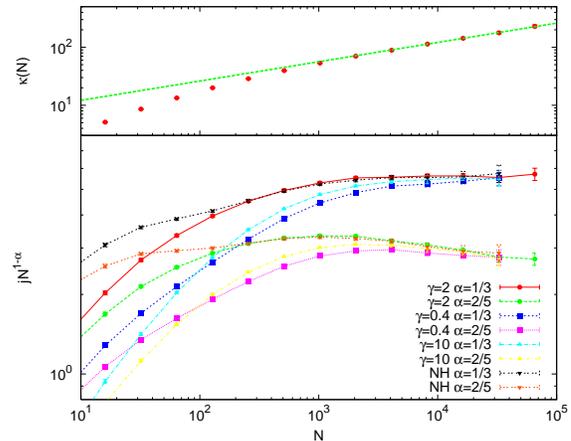}
\caption{(Color online) Heat current as a function of $N.$ Root
mean square errors from $O(10^5-10^6)$ measurements are shown except
when they are smaller than the points.  (Top) Conductivity versus
$N$.  The last five points fit to a slope of $\alpha =0.333\pm
0.004.$ The baths are described in Fig.(\ref{fig:temp}).  (Bottom)
$jN^{1-\alpha}$ versus $N$ for $\alpha=1/3$ and $\alpha=2/5$.  In
the large $N$ regime, $\alpha$ is definitely less than 2/5 and
appears to agree quite well with the 1/3 prediction for all data
sets.  Langevin baths with $\gamma$=0.4, 2, and 10, and one data
set with Nose-Hoover baths, are shown. }
\label{fig:conduct}
\end{center}
\end{figure}

The heat current flowing in steady state is measured as a function
of $N$ for equal mass FPU chains.  The time averaged current $j$
is defined by $j = -\langle\sum_i \dot{x}_iV^\prime(x_{i+1}-x_i)/N\rangle$,
where $x_i$ is the displacement from equilibrium of the $i^{th}$
particle. As shown in Fig.~\ref{fig:conduct}, $\kappa(N)=-j N/\Delta
T$ satisfies Eq.(\ref{defalpha}) for large $N$ with $\alpha =
0.333\pm 0.004,$ in strong agreement with the RG prediction.

To test the sensitivity of this result to different baths, we run
simulations with Langevin baths with different damping constants
$\gamma$=0.4, 2, and 10. We also replace the stochastic Langevin
baths with deterministic Nose-Hoover~\cite{NoseHoover} thermostats,
for which we use the fourth order Runge-Kutta integrator.
Fig.(\ref{fig:conduct}) compares the RG prediction and the MCT
prediction for systems with these different baths and bath parameters.
As can be seen in the figure, an asymptotic exponent of 1/3 is
attained for {\it all\/} these systems, whereas the apparent exponents
for smaller $N$ depend on system parameters.  Moreover, it is
possible to understand the deviation of the apparent exponent from
1/3 for small system sizes.  As shown in Ref.\cite{LLPreview}, if
the damping constant for the Langevin baths is very large or small,
there is a large `contact resistance' at the boundaries of the
chain. The current only depends weakly on $N,$ resulting in an
apparent $\alpha > 1/3.$ (Similar considerations apply to Nose-Hoover
baths~\cite{foot:NHresponse}.) This is confirmed by our results:
the plot for $\gamma=2$ reaches the asymptotic limit fastest, whereas
$\gamma = 0.4,10$ have apparent exponents closer to 0.4 for small
$N.$ Taken together, the large-$N$ exponent of 1/3, the universality
with respect to bath parameters, and the explanation for how the
apparent exponent behaves as a function of $N$ and $\gamma$ for
small $N$ convincingly supports the RG prediction.

Since Fourier's law is only applicable in the linear response regime,
we halve the temperature difference between the ends and simulate
the same spring system with Langevin baths with $\gamma=2$, $T_L=1.625$
and $T_R=0.875$. Though the error bars are larger, $\alpha$ still
agrees with 1/3, verifying that the system is in the linear response
regime.

The exponent $\alpha$ measured in our simulations of FPU chains
clearly differs from the measurements from other
simulations~\cite{LLP98,LLP03,DLLP1,WangLi}. This disagreement is
mainly because very large system sizes are needed. Moreover, we use
a step size $h=0.0025-0.005$ that is an order of magnitude smaller
than the step size used for the dynamics in~\cite{LLP03}; by comparing
numerical and exact results for harmonic springs, we have found
that a small $h$ is necessary for proper convergence. Finally, we
have shown the results as a function of the coupling to the bath,
which has not been done before, and explained why the baths increase
the apparent $\alpha,$ especially if the coupling is not optimal.
We note that only equilibrium correlations for a periodic FPU-$\beta$
chain are shown in Ref.~\cite{DLLP1}. As discussed in the second
paper in Ref.~\cite{RC}, this is problematic.

However, in Ref.~\cite{LLP03}, where purely quartic springs ($k_2=0$)
and system sizes similar to our simulations were used, $\alpha \sim
0.44$ was extrapolated from an indirect measurement and $\alpha
\gsim 0.4$ was measured directly from nonequilibrium simulations.
We have found $\alpha = 0.38-0.39$ for this system.  Because the
analytical results all obtain the same asymptotic $\alpha$ for the
FPU-$\beta$ and quartic models, and we have shown that the apparent
$\alpha$ decreases with $N$ for the FPU-$\beta$ model, we believe
that $\alpha$ for the quartic system will also asymptotically change
to $1/3.$ The alternative, that $\alpha$ for the FPU-$\beta$ chain
will reverse its change with $N$ and revert to $\approx 0.4,$ seems
unlikely. Indeed, the latest MCT results~\cite{DLLP2} obtain $\alpha
= 0.5$ for even potentials (including the quartic model and the
FPU-$\beta$ model) which is quite far from earlier~\cite{DLLP1} and
our numerical results.  Nevertheless, it is unclear why the pure
quartic system should need exceptionally large $N.$ A final resolution
of the issue requires an analytical demonstration of an error in
one of the competing methods.

The existence of a paradoxical LT on small scales for slight
alterations to the equal mass FPU chain could be partly responsible
for the unusually large $N$ needed to reach the asymptotic limit
(most notably for purely quartic potentials).  Conversely, slight
changes can improve the convergence to $\alpha=1/3$ as seen by
adding a cubic term~\cite{DLLP1,DLLP2}, transverse motion~\cite{WangLi},
or collisions~\cite{tunable,LLP03}.  Recent simulations of nanotubes
obtain $\alpha = 1/3$~\cite{nanosim}.

In summary, we have shown that the local temperature behaves
paradoxically when an $O(1/N)$ temperature gradient is applied to
FPU chains with unequal masses; coarse-graining with any $O(1)$
averaging length does not cure this.  This renders existing analytical
models for heat conductivity inapplicable.  However, a well-behaved
LT is established for equal mass chains, or when the baths are at
the same temperature~\cite{bergmann55}.  For equal mass chains,
large scale simulations support the much debated heat conductivity
exponent being 1/3 as predicted~\cite{RG}.

We thank Sriram Ramaswamy, Ram Ramaswamy, and Peter Young for useful
comments.


\begin{thebibliography}{}
\bibitem{foot:NHresponse} The inverse response time for Nose-Hoover
thermostats alters the contact resistance, analogous to the Langevin
damping constant~\cite{LLPreview}.  We use a unit response time.
\end{thebibliography}
\end{document}